\documentclass[aps,pra,twocolumn,showpacs,superscriptaddress,floatfix]{revtex4}
\usepackage{graphicx}
\usepackage{nicefrac}
\usepackage{amsmath}
\usepackage{amsfonts}
\usepackage{amssymb}
\usepackage{amsthm}
\usepackage{epsf}
\usepackage{bm}
\usepackage{bbm}
\usepackage{longtable}

\usepackage{dcolumn}

\newcolumntype{.}{D{x}{}{-1}}

\newcommand{\balpha}{\bm{\alpha}}

\newcommand{\bfr}{\bm{r}}
\newcommand{\bfk}{\bm{k}}
\newcommand{\bfp}{\bm{p}}

\newcommand{\bfx}{\bm{x}}

\newcommand{\hp}{\hat{\bfp}}

\newcommand{\vare}{\varepsilon}

\newcommand{\pr}{^{\prime}}

\newcommand{\SixJ}[6]{
        \left\{
        \begin{array}{ccc}
        #1  & #2  & #3 \\
        #4  & #5  & #6 \\
        \end{array}
        \right\}
        }

\newcommand{\lbr}{\left<} \newcommand{\rbr}{\right>}

\begin{document}

\title{The off-resonant dielectronic recombination
in a collision of an electron with a heavy hydrogen-like ion}

\author{Vladimir A. Yerokhin}
\affiliation{Institute of Physics, University of Heidelberg, Philosophenweg
  12, D-69120 Heidelberg, Germany}
\affiliation{Gesellschaft f\"ur Schwerionenforschung, Planckstra{\ss}e 1,
D-64291 Darmstadt, Germany}
\affiliation{Center for Advanced Studies, St.~Petersburg State
Polytechnical University, Polytekhnicheskaya 29, 
St.~Petersburg 195251, Russia}

\author{Andrey Surzhykov}
\affiliation{Institute of Physics, University of Heidelberg, Philosophenweg
  12, D-69120 Heidelberg, Germany}
\affiliation{Gesellschaft f\"ur Schwerionenforschung, Planckstra{\ss}e 1,
D-64291 Darmstadt, Germany}

\begin{abstract}

The recombination of an electron with an (initially) hydrogen-like ion
is investigated. The effect of the electron-electron interaction is
treated rigorously to the first order in the parameter $1/Z$ and within
the screening-potential approximation to higher orders in $1/Z$, with $Z$
being the nuclear charge number. The two-electron correction contains the
dielectronic-recombination part, which contributes to the
process not only under the resonance condition for the projectile energy but
also in the regions far from resonances. The mechanism of
the off-resonant dielectronic recombination is studied in detail.

\end{abstract}


\maketitle

%%%%%%%%%%%%%%%%%%%%%%%%%%%%%%%%%%%%%%%%%%%%%%%%%%%%%%%%%%%%%%%%%%%%%%%%%%%
\section{Introduction}

One of the main processes occurring in collisions of a highly charged ion with
an electron is radiative recombination (RR), in which the electron is captured
from the continuum into a bound state with emission of a photon.
In the case when the ion initially
possesses one or several electrons, the electron capture can proceed also via
dielectronic recombination (DR), in which the energy excess goes to the
excitation of the second electron, which then returns to the ground state
via a radiative decay. DR is a resonant process and is
usually studied under the condition that the excess energy is very close to the
excitation energy of the second electron. In this case, DR is the
dominant recombination channel, whereas RR is responsible for a nonresonant
background. Outside of the resonance region, RR is the dominant process.

In the zeroth approximation, RR and DR are often considered as
two independent recombination channels, which can be calculated separately and
combined additively \cite{spies:92}. More accurate calculations
include the effects of quantum interference between DR and RR
\cite{zimmermann:97,tokman:02,mohamed:02}. Generally speaking, at the level of
precision where effects of the electron-electron interaction come
into play, RR and DR cannot be meaningfully separated. Outside of the
resonance region, DR is essentially a correction to RR due to the
electron-electron interaction and induces a contribution of the same order
of magnitude as other two-body effects, e.g.,
the screening of the nuclear charge by core electron(s).

The accuracy of experimental investigations of the RR process with heavy highly
charged ions has gradually increased during the past years
\cite{stoehlker:94,stoehlker:97:prl,stoehlker:99:prl}, reaching the level
on which the electron-electron interaction effect can be clearly identified
\cite{reuschl:08}. A disagreement with the one-electron theory observed
in the state-selective study of RR into hydrogen-like uranium
\cite{reuschl:08} calls for an accurate theoretical description of the
electron-electron interaction effect.

Most of the previous calculations of the RR process into heavy few-electron ions
accounted for the electron correlation by means of the
Dirac-Fock method \cite{fritzsche:05,trzhaskovskaya:03}, disregarding the
off-resonant DR mechanism. Evidences that the omitted contribution might be
significant were reported in Refs.~\cite{korol:06,korol:04}, where
a part of the off-resonant DR (involving photon emission from a core electron)
was studied. It was claimed that, for many-electron systems, this mechanism can
significantly influence the RR process, yielding an order-of-magnitude
enhancement in some specific cases.

In the present investigation we perform an {\em ab initio}
calculation of the electron-electron interaction
effect on RR into a heavy hydrogen-like ion in the non-resonant
region of energies of the incoming electron.
A particular emphasis will be made on the contribution of the
off-resonant DR, as this effect has not been carefully studied before.
A similar study of RR into a helium-like uranium have been reported
previously in Ref.~\cite{yerokhin:00:recpra}.

Relativistic units ($\hbar = c = 1$) are used in this paper.

%%%%%%%%%%%%%%%%%%%%%%%%%%%%%%%%%%%%%%%%%%%%%%%%%%%%%%%%%%%%%%%%%%%%%%%%%%%
\section{General approach}

We consider RR of an electron with
an (initially) hydrogen-like ion. The initial state consists of
the incident electron with the asymptotic momentum $\bfp$, the energy
$\vare = \sqrt{\bfp^2+m^2}$, and the spin projection $\mu_s = \pm
1/2$ and the bound (core) electron in the
state $a'$ with the relativistic angular quantum number $\kappa_{a'}$ and
the momentum projection $\mu_{a'}$.
In the final state, there is the two-electron
bound state with the angular momentum $J$ and the projection $M$ and the outcoming
photon with the momentum $\bfk$, and the energy $\omega = |\bfk|$.
The wave function of the final two-electron state is
\begin{align} \label{000}
|JM\bigr> = N \sum_{\mu_a\mu_v}
 C^{JM}_{j_a \mu_a\,j_v\mu_v} |\kappa_a\mu_a\bigr>|\kappa_v\mu_v\bigr>\,,
\end{align}
where $a$ and $v$ stand for the core and the valence electron, respectively,
and $N = 1/\sqrt{2}$ for the equivalent electrons and $N=1$ otherwise. The
core electron state is not changed in the process, thus $\kappa_a = \kappa_{a'}$.
The wave function (\ref{000}) is not antisymmetrized since we choose
to perform antisymmetrization explicitly for the amplitude.

General formulas are conveniently written in the center-of-mass frame, which
practically coincides with the rest system of the ion.
The direction of the $z$ axis of the coordinate system is chosen to
be the direction of the emitted photon.

In the following, we will assume that the fine-structure levels with different
$J$'s are {\em not} resolved in the experiment (as is the case for the experiments
conducted so far).

%%%%%%%%%%%%%%%%%%%%%%%%%%%
\subsection{Zeroth order}

To the zeroth order, we neglect the electron-electron interaction. The
core electron does not participate in the process and
the transition amplitude is written as
\begin{align} \label{eq8}
\tau^{(0)}_{\mu_s\mu_{a'}JM} = N \sum_{\mu_a\mu_v} C^{JM}_{j_a \mu_a\,j_v\mu_v}\,
\delta_{\mu_{a}\mu_{a'}}\, \tau^{(0)}_{\mu_s\mu_v}\,,
\end{align}
where $\tau^{(0)}_{\mu_s\mu_v}$ is the amplitude for the recombination
with the bare nucleus.
It reads
\cite{eichler:95:book}
\begin{equation}\label{eq3}
  \tau^{(0)}_{\mu_s\mu_v} = \lbr v|\balpha\cdot \hat{\bm u}^*e^{-i\bfk\cdot\bfr} |p\rbr \,,
\end{equation}
where $|v\bigr> \equiv |\kappa_v\mu_v\bigr>$ denotes the bound state,
$|p\bigr> \equiv |{\bm p}\mu_s\bigr>$ is the Dirac continuum-state wave
function with a definite asymptotic momentum, and $\hat{\bm u}$ is the unit
polarization vector of the emitted photon.
After summation over the final states and
averaging over the initial states, the differential cross section
of the process is written as
\begin{align}
 \frac{d\sigma^{(0)}}{d\Omega} &\ =
   \frac1{2j_{a}+1}\, \frac{\alpha\omega m}{4\beta^2 \vare^2}\,
   \sum_{\mu_s\mu_{a'}JM} \bigl| \tau^{(0)}_{\mu_s\mu_{a'}JM}\bigr|^2
\nonumber  \\ &
 = N^2\,\frac{\alpha\omega m}{4\beta^2 \vare^2}\,
    \sum_{\mu_s\mu_v} \bigl| \tau^{(0)}_{\mu_s\mu_v}\bigr|^2\,,
 \label{eq4b}
\end{align}
where $\beta = \sqrt{1-m^2/\vare^2}$. Because of the summation over the
initial and final states, the cross section does not depend on the
polarization of the emitted photon, which can be fixed arbitrary.
The formula (\ref{eq4b}) differs from the corresponding expression
for the RR into the bare nucleus \cite{eichler:95:book}
only by a factor of $N^2$ ($ = 1/2$ for the recombination into the ground
state and $1$ otherwise).

The energy of the emitted photon in Eq.~(\ref{eq4b})
is fixed by the energy conservation
condition $\omega = \vare-\vare_v$ or, more generally,
$\omega = \vare-m+\vare_{\rm io}$, where $\vare_{\rm io}$ is the ionization
energy of the atom in the final state.

%%%%%%%%%%%%%%%%%%%%%%%%%%%
\subsection{Electron-electron interaction}

\begin{figure*}
  \centerline{\includegraphics[width=0.8\textwidth]{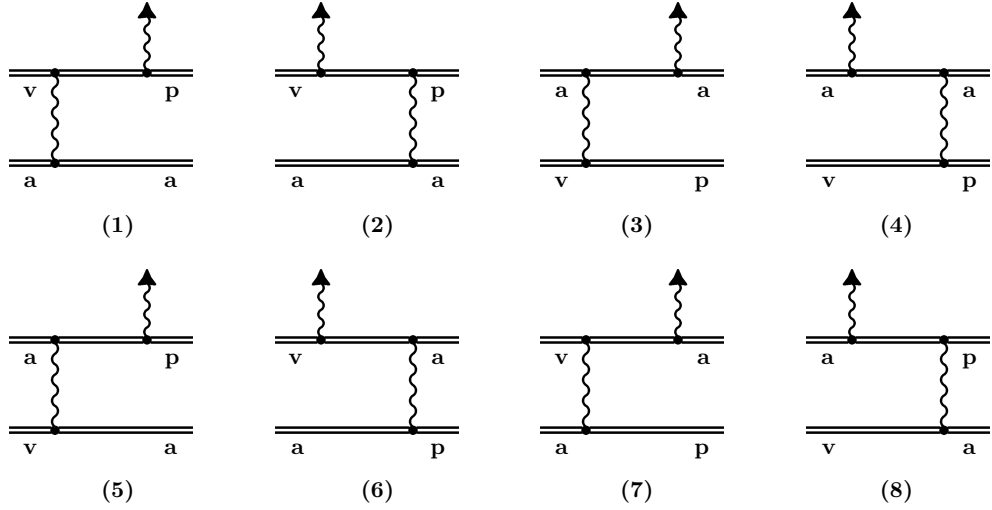}}
\caption{The one-photon exchange correction to the transition amplitude of the
radiative recombination of an electron with a hydrogen-like ion.
The double line indicates an electron propagating in the field of a nucleus.
The wavy line with an arrow denotes the emitted photon. The incoming
electron is denoted as $p$, $a$ is the initially bound (core) electron, and $v$
is the captured (valence) electron.
  \label{fig:1} }
\end{figure*}

For a heavy few-electron ion, the electron-electron interaction can be
effectively accounted for by a perturbative expansion in the parameter
$1/Z$. The first-order correction is induced by a single virtual-photon exchange
between the electrons, shown diagrammatically in Fig.~\ref{fig:1}.
The corresponding correction to the differential cross section can be written
as
\begin{align} \label{eq10}
\frac{d\sigma^{(1)}}{d\Omega} &=
\delta_{\omega} \frac{d\sigma^{(0)}}{d\Omega}
\nonumber \\ &
+
   \frac1{2j_{a}+1}\,\frac{\alpha\omega m}{4\beta^2 \vare^2}\,
   \sum_{\mu_s\mu_{a'}JM} 2\,{\mathrm{Re}} \left[ \tau^{{(0)}^*}_{\mu_s\mu_{a'}JM}
      \tau^{{(1)}}_{\mu_s\mu_{a'}JM}\right]\,,
\end{align}
where $\tau^{{(1)}}$ denotes the first-order correction to the amplitude and
$\delta_{\omega}$ is induced by the change of the energy
of the emitted photon (because of the shift of the energy of the final state
due to the presence of the second electron),
\begin{align}
\delta_{\omega} \frac{d\sigma^{(0)}}{d\Omega} =
     \left.\frac{d\sigma^{(0)}}{d\Omega}\right|_{\omega=\omega^{(0)}+\delta\omega}
     -\left.\frac{d\sigma^{(0)}}{d\Omega}\right|_{\omega=\omega^{(0)}}\,.
\end{align}
We note that Eq.~(\ref{eq10}) assumes that the perturbative regime
($|\tau^{(1)}| \ll |\tau^{(0)}|$) takes place.

%%%%%%%%%%%%%%%%%%%%%%%%%%%%%%%%%%%%%%%%%%%%%%%%%%%%%%%%%%%%%%%%%%%%%%%%%%%%%%%%%%
%
% Fig. 1
%
%%%%%%%%%%%%%%%%%%%%%%%%%%%%%%%%%%%%%%%%%%%%%%%%%%%%%%%%%%%%%%%%%%%%%%%%%%%%%%%%%%

Since the fine-structure sublevels of the final state are not
resolved in the experiment, the dependence on $J$ and $M$ can be eliminated
already in the general formulas. To achieve this, we write the
correction to the amplitude as
\begin{align}
\tau^{(1)}_{\mu_s\mu_{a'}JM} = N \sum_{\mu_a\mu_v} C^{JM}_{j_a \mu_a\,j_v\mu_v}
   \tau^{(1)}_{\mu_s\mu_{a'}\mu_v\mu_a}\,,
\end{align}
where $\tau^{(1)}_{\mu_s\mu_{a'}\mu_v\mu_a}$ does not depend on $J$ and $M$.
Inserting this formula
and Eq.~(\ref{eq8}) into Eq.~(\ref{eq10}) and performing
summations, we obtain
\begin{align} \label{eq12}
\frac{d\sigma^{(1)}}{d\Omega} =
\delta_{\omega} \frac{d\sigma^{(0)}}{d\Omega}
+
   \frac{N^2}{2j_{a}+1}\,\frac{\alpha\omega m}{4\beta^2 \vare^2}\,
   \sum_{\mu_s\mu_{v}} 2\,\mathrm{Re} \left[ \tau^{{(0)}^*}_{\mu_s\mu_{v}}
      \tau^{{(1)}}_{\mu_s\mu_{v}}\right]\,.
\end{align}
Here,
\begin{align}
\tau^{(1)}_{\mu_s\mu_{v}} \equiv \sum_{\mu_a} \tau^{(1)}_{\mu_s\mu_a\mu_v\mu_a}\,,
\end{align}
is the amplitude of the recombination with a closed-shell atom. So, we obtain
that, in the situation when the fine-structure levels are not resolved in the
experiment, formulas for the RR with a hydrogen-like ion differ from those
for the RR with a helium-like ion only by a prefactor of
$N^2/(2j_a+1)$.

General expressions for the one-photon exchange correction to the RR
of an electron with a heavy ion were derived in
Ref.~\cite{shabaev:02:rep}.
(For the closed-shell ions, such derivation was reported also in
Ref.~\cite{yerokhin:00:recpra}). The
correction to the transition amplitude consists of 8 terms corresponding to
the 8 diagrams on Fig.~\ref{fig:1},
\begin{align}
\tau^{(1)}_{\mu_s\mu_{v}} = \sum_{i=1}^8 \tau^{(1,i)}\,.
\end{align}
The individual contributions for each diagram are given by
\begin{align} \label{tau1}
\tau^{(1,1)} =
\sum_{\mu_a}
  \sum_{n\ne v}
        \frac{\bigl< va|I(0)| na \bigr> \bigl< n|\balpha\cdot \hat{\bm u}^*e^{-i\bfk\cdot\bfr}|p\bigr>}
    {\vare_v-\vare_n}  \,,
\end{align}
\begin{align} \label{tau2}
\tau^{(1,2)} =
\sum_{\mu_a}
  \sum_{n}
        \frac{\bigl< v|\balpha\cdot \hat{\bm u}^*e^{-i\bfk\cdot\bfr}|n\bigr> \bigl< na|I(0)| pa\bigr>}
    {\vare-\vare_n(1-i0)} \, ,
\end{align}
\begin{align} \label{tau3}
\tau^{(1,3)}=
\sum_{\mu_a}
\sum_{n}
        \frac{\bigl< va|I(\vare -\vare_v)| pn\bigr> \bigl< n|\balpha\cdot \hat{\bm u}^*e^{-i\bfk\cdot\bfr}|a\bigr>}
    {\vare_a+\vare_v-\vare-\vare_n(1-i0)} \,,
\end{align}
\begin{align} \label{tau4}
\tau^{(1,4)}=
\sum_{\mu_a}
\sum_{n}
        \frac{\bigl< a|\balpha\cdot \hat{\bm u}^*e^{-i\bfk\cdot\bfr}|n\bigr> \bigl< vn|I(\vare-\vare_v)| pa\bigr>}
    {\vare_a-\vare_v+\vare-\vare_n(1-i0)} \,,
\end{align}

\begin{align} \label{tau5}
\tau^{(1,5)}=&\
-\sum_{\mu_a}
  \sum_{n\ne v}
        \frac{\bigl< av|I(\vare_v-\vare_a)| na \bigr> \bigl< n|\balpha\cdot \hat{\bm u}^*e^{-i\bfk\cdot\bfr}|p\bigr>}
    {\vare_v-\vare_n}
\nonumber \\ &
   -\frac12\, \sum_{\mu_a\mu_{v'}}
\bigl< av|I\pr(\vare_v-\vare_a)|v'a\bigr>
    \bigl< v'| \balpha\cdot \hat{\bm u}^*e^{-i\bfk\cdot\bfr} |p\bigr> \,,
\end{align}
\begin{align} \label{tau6}
\tau^{(1,6)}=
-\sum_{\mu_a}
  \sum_{n}
        \frac{\bigl< v|\balpha\cdot \hat{\bm u}^*e^{-i\bfk\cdot\bfr}|n\bigr> \bigl< na|I(\vare-\vare_a)| ap\bigr>}
    {\vare-\vare_n(1-i0)} \, ,
\end{align}
\begin{align} \label{tau7}
\tau^{(1,7)}=
-\sum_{\mu_a}
\sum_{n}
        \frac{\bigl<av|I(\vare-\vare_a)| pn\bigr> \bigl< n|\balpha\cdot \hat{\bm u}^*e^{-i\bfk\cdot\bfr}|a\bigr>}
    {\vare_a+\vare_v-\vare-\vare_n(1-i0)} \,,
\end{align}
\begin{align}
 \label{tau8}
\tau^{(1,8)}=
-\sum_{\mu_a}
\sum_{n}
        \frac{\bigl< a|\balpha\cdot \hat{\bm u}^*e^{-i\bfk\cdot\bfr}|n\bigr> \bigl< vn|I(\vare_a-\vare_v)|ap\bigr>}
    {\vare_a-\vare_v+\vare-\vare_n(1-i0)} \,.
\end{align}
Here, $I(\Delta)$ is the operator of the electron-electron interaction,
\begin{align}
I(\Delta) = e^2 \alpha_{\mu}\alpha_{\nu}\,D^{\mu\nu}(\Delta,\bfx_{12})\,,
\end{align}
where $D^{\mu\nu}$ is the photon propagator. In the Feynman gauge, the
operator $I$ takes the form
\begin{align}
I(\Delta) = \frac{\alpha}{4\pi}\,
\frac{1-\balpha_1\cdot\balpha_2}{x_{12}}\,e^{i|\Delta|x_{12}}\,.
\end{align}
The summations over $n$ in Eqs.~(\ref{tau1})-(\ref{tau8})
extend over the complete spectrum of the Dirac
equation. The second term on the right-hand-side of Eq.~(\ref{tau5})
corresponds to the $n=v$ contribution excluded from the summation in the first
term. The prime on the operator $I$ denotes the
derivative with respect to the energy argument. The state $v'$ is the $n=v$
state with the angular momentum projection $\mu_{v'}$. The small imaginary addition to
the intermediate-state energies in the energy denominators fixes the position of
the energy argument of the electron propagator $G({\cal E})$ with respect to the
branch cuts for $|{\cal E}|>m$.

We now turn to the physical interpretation of individual diagrams in
Fig.~\ref{fig:1}. The first two graphs represent the effect of the screening of
the nuclear charge by the core electron. The
corresponding corrections [$\tau^{(1,1)}$ and $\tau^{(1,2)}$]
can be regarded as the first-order perturbations of the zeroth-order
amplitude (\ref{eq3}) by the screening potential of the core electron,
\begin{align} \label{vscr}
V_{\rm scr}(x) = \alpha \int_0^{\infty}dy\,y^2\,\frac1{\mathrm{max}(x,y)}\,
 \left[ g_a^2(y)+f_a^2(y)\right]\,,
\end{align}
where $g_a$ and $f_a$ are the upper and the lower radial components of the
core electron state. The screening effect can easily be accounted for
to all orders in $1/Z$ by evaluating the zeroth-order amplitude for an electron in
a combination of the nuclear and the screening potentials. Such treatment is
exactly equivalent to the frozen-core Dirac-Fock method (as
the core in our case contains one electron only).

The contribution of diagram (5) in Fig.~\ref{fig:1} can be interpreted to
represent the electron correlation on the bound-electron wave function (also
known as the ``relaxation'' effect). It can be partly included by the standard
many-body techniques like many-body perturbation theory or the
multiconfiguration Dirac-Fock method.

The contribution of diagrams (4) and (8) in Fig.~\ref{fig:1}
contain resonant parts that become prominent
when the projectile energy approaches the region where $\vare \approx
\vare_v-\vare_a+\vare_n > m$, with $\vare_n$ being a Dirac bound-state energy.
When the resonance condition is fulfilled, the core electron
gets excited into a higher-lying bound state, which corresponds to the
standard resonant DR mechanism. In that case, the electron
propagator can be replaced by a contribution of the single state responsible for the
resonance (the so-called ``resonance'' approximation), thus greatly
simplifying the problem. In the region far from the resonance,
however, the core
electron gets ``excited'' in all possible virtual states of the energy spectrum,
so that the usage of the full Dirac propagator becomes essential in the
description of this process.

The diagrams (3), (6) and (7) in Fig.~\ref{fig:1} correspond to other
processes with participation of the core electron,
in which the full energy spectrum of virtual states is probed. 
We will refer to the contribution of all the diagrams (3),
(4), (6), (7), and (8) as the (off-resonant) DR correction. 
So, in the present work, the term DR is used to refer to the recombination
with an assistance of the second electron, rather than only to the resonant part
of this process, as is customary. It should be noted that the separation of the total
two-electron effect in several parts is
to a large extent artificial (e.g., the DR
correction defined in this way is not gauge invariant). Its main justification
is that the screening and correlation parts are easily accounted for by
standard methods, whereas the DR part is not. The sum of all two-electron
contributions, however, is gauge invariant and derived rigorously within QED.

So, we represent the total RR cross section as a sum of four terms,
\begin{align} 
\sigma = \sigma^{(0)}+ \sigma_{\rm scr}+ \sigma^{(1)}_{\rm corr}+
\sigma^{(1)}_{\rm DR}\,,
\end{align}
where $\sigma^{(0)}$ is the zeroth-order cross section,
$\sigma_{\rm scr}$ is the correction induced by the screening potential
$V_{\rm scr}$ included to all orders,
$\sigma^{(1)}_{\rm corr}$ is the correlation correction
induced by $\tau^{(1,5)}$, and
$\sigma^{(1)}_{\rm DR}$ is the off-resonant DR contribution induced by $\tau^{(1,3)}$, $\tau^{(1,4)}$,
$\tau^{(1,6)}$, $\tau^{(1,7)}$, and $\tau^{(1,8)}$. The screening correction
is calculated with the ``correct'' energy of the emitted photon
and thus includes the $\delta_{\omega}$ correction in Eq.~(\ref{eq10}).
We assume that the projectile energy is far enough
from the resonance condition
to ensure that the perturbative regime is valid.

%%%%%%%%%%%%%%%%%%%%%%%%%%%%%%%%%%%%%%%%%%%%%%%%%%%%%%%%%%%%%%%%%%%%%%%%%%%%%%%%%%
%
%%%%%%%%%%%%%%%%%%%%%%%%%%%%%%%%%%%%%%%%%%%%%%%%%%%%%%%%%%%%%%%%%%%%%%%%%%%%%%%%%%
\section{Numerical evaluation}

The integration over angular variables in the general formulas
of the previous section can be
performed by means of the standard Racah algebra, as illustrated in
Ref.~\cite{yerokhin:00:recpra}. The resulting formulas for the zeroth-order
transition amplitude and for the first-order corrections are given in Appendix.
Performing our calculations, we found several sign mistakes in
Ref.~\cite{yerokhin:00:recpra}. Namely, the contributions of Eqs.~(\ref{F7})
and (\ref{F8}) were accounted for with the opposite sign in that
work. Moreover, the incorrect sign was present in the first term of
Eq.~(\ref{F5}) in the case of the capture into the $2p_{1/2}$ state.

The zeroth-order cross section  $\sigma^{(0)}$ and the screening correction
$\sigma_{\rm scr}$ were evaluated according to Eqs.~(\ref{eq4b}) and
(\ref{eq5}). The radial bound and continuum wave functions were obtained by
solving the Dirac equation with an extended-nucleus Coulomb potential and the
screening potential of the core electron, by using
the RADIAL package by Salvat {\em et al.} \cite{salvat:95:cpc}.

The calculation of the first-order corrections $\sigma^{(1)}_{\rm corr}$
and $\sigma^{(1)}_{\rm DR}$ was more complicated, due to a larger number of
radial integrations and the summations over the complete spectrum of the Dirac
equation. In the evaluation of the $\tau^{(1,5)}$ correction, we employed the
dual kinetically balanced $B$-spline basis set \cite{shabaev:04:DKB} to
represent the Dirac spectrum. In most of other cases, we used the analytical
representation of the radial Dirac Coulomb Green function in terms of
Whittaker functions \cite{mohr:98}. For simplicity, we used the point-nucleus
Green function, since the effect of the finite nuclear size turned out
to be negligibly small. In the evaluation of the $\tau^{(1,4)}$ and
$\tau^{(1,8)}$ corrections, we used the finite basis set when the energy
argument of the Green function was smaller than the electron rest mass
${\cal E} < m$, and the exact Green function, otherwise. The Dirac Coulomb Green
function with ${\cal E} > m$ is a complex-valued function and a care must be
taken in order to choose the appropriate branch of it.
The sign of the imaginary part of the Green function
is fixed by the sign of the small imaginary addition in the energy
denominators of Eqs.~(\ref{tau2})-(\ref{tau7}) and discussed in detail in
Ref.~\cite{yerokhin:00:recpra}.

A problem emerges in the numerical evaluation of the radial integrals when they
contain, apart from the Bessel function, two continuum-state wave
functions. In this case, the integrand is a rapidly oscillating function that
falls off very slowly at large radial distances. In our case, such situation
arises only in the evaluation of the $\tau^{(1,8)}$ correction for projectile
energies $\vare > m-\vare_a+\vare_v$. [The problem appears also for the
$\tau^{(1,2)}$ correction if it is evaluated perturbatively but not if it is
evaluated to all orders.]

Our scheme of evaluation of radial integrals was as follows. First, we
introduce the parameter $R_1$ that
represents the distance at which all bound-state wave functions become
negligibly small. (Typically, $R_1 = 80/Z$ a.u.) At the distances
$r>R_1$, all radial integrals with bound-state wave functions reach their
asymptotic values, so that the problem reduces to the evaluation of
one-dimensional integrals of the form
\begin{align}
 \int_{R_1}^{\infty}dr\,r^2\,j_l(\omega r)\,f^i(r)\,\phi_{\infty}^{j}(r)\,,
\end{align}
where $j_l$ is a spherical Bessel function,
$f^i$ is a radial component of the continuum-state Dirac wave function
and $\phi_{\infty}^{j}$ is the irregular solution of the Dirac equation
(originating from the Green function). To evaluate these integrals,
we introduce a small regulator parameter $\alpha>0$ and multiply the
integrand by $\exp(-\alpha r)$. The regularized integrals are cut off at large
distances by a parameter $R_2 \propto 1/\alpha$ and evaluated
numerically with Gauss-Legendre quadratures. The typical value of the
regulator was $\alpha = 10^{-3}$. We checked that decreasing the
regulator by a factor of 10 does not influence our numerical results
significantly.

%%%%%%%%%%%%%%%%%%%%%%%%%%%%%%%%%%%%%%%%%%%%%%%%%%%%%%%%%%%%%%%%%%%%%%%%%%%%%%%%%%
%
% Fig. 2
%
\begin{figure*}[thb]
  \centerline{\includegraphics[width=\textwidth]{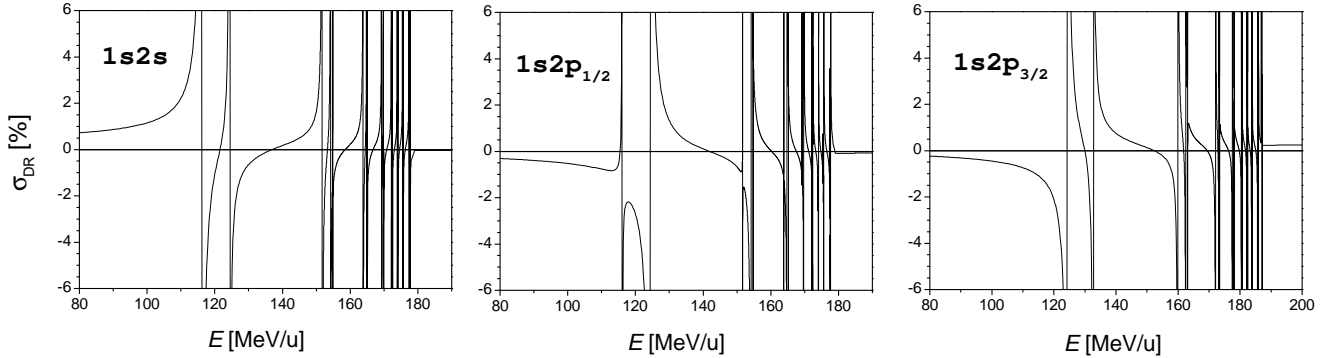}}
\caption{
The contribution of dielectronic recombination $\sigma_{\rm DR}^{(1)}$
for the capture into the $1s2s$, $1s2p_{1/2}$, and $1s2p_{3/2}$ states of the initially
hydrogen-like uranium
as a function of the projectile energy $E$,
in units per cent of the
zeroth-order cross section $\sigma^{(0)}$.
The threshold energies of the resonant
dielectronic recombination for the capture into the $1s2s$, $1s2p_{1/2}$, and
$1s2p_{3/2}$ states  are $E_{0} = 178.6$,
$178.4$, and $186.7$~MeV/u, respectively.
  \label{fig:2} }
\end{figure*}
%%%%%%%%%%%%%%%%%%%%%%%%%%%%%%%%%%%%%%%%%%%%%%%%%%%%%%%%%%%%%%%%%%%%%%%%%%%%%%%%%%

%%%%%%%%%%%%%%%%%%%%%%%%%%%%%%%%%%%%%%%%%%%%%%%%%%%%%%%%%%%%%%%%%%%%%%%%%%%%%%%%%%
%
% Fig. 3
%
\begin{figure*}[thb]
  \centerline{\includegraphics[width=0.95\textwidth]{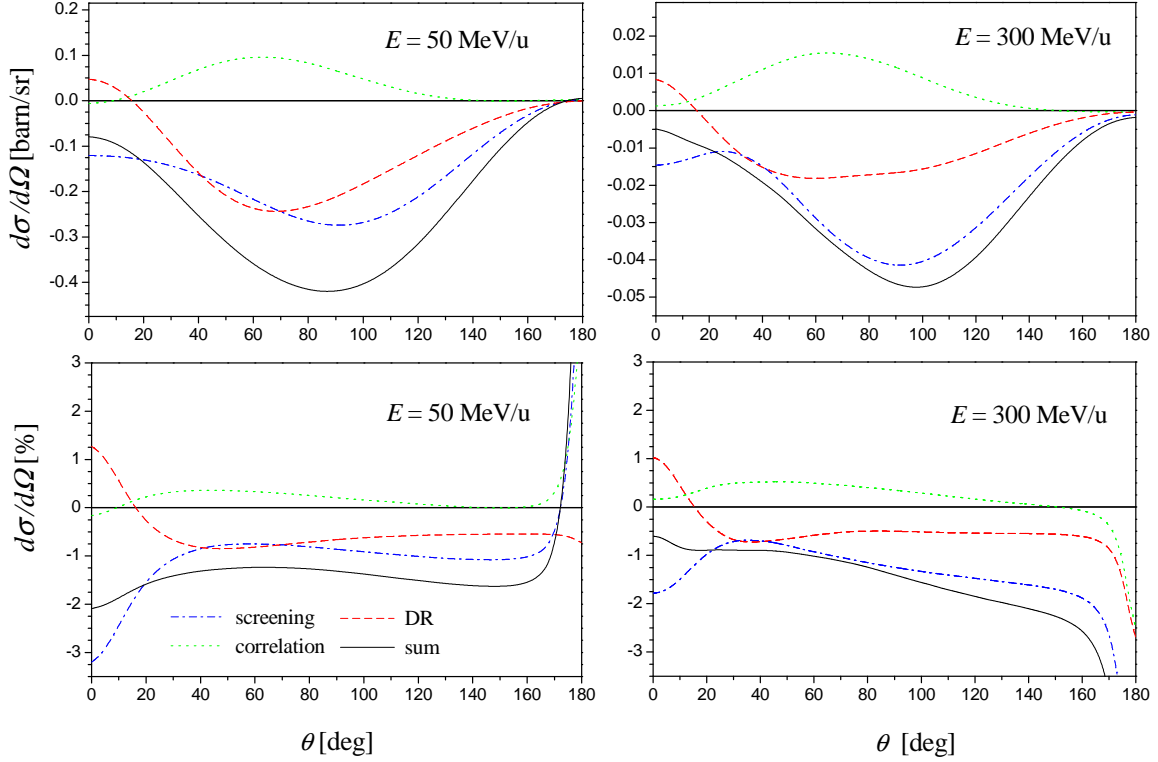}}
\caption{
Individual two-electron contributions to the differential cross section of RR
into the ground state of the initially hydrogen-like uranium, for two values
of the projectile energy, $E = 50$ MeV/u (left column) and 300 MeV/u (right
column), as a function of the observation angle $\theta$. 
The upper graphs represent the absolute contributions to the cross
section in barns/sr and the lower graphs, the relative magnitude of
the corrections, in units per cent of the zeroth-order cross section
$d\sigma^{(0)}$. The dash-doted line (blue on-line) corresponds to the
screening part; the dotted line (green on-line), to the correlation
correction; the dashed line (red on-line), to the DR correction; and the solid
line (black on-line), to the total two-electron effect. 
  \label{fig:3} }
\end{figure*}
%%%%%%%%%%%%%%%%%%%%%%%%%%%%%%%%%%%%%%%%%%%%%%%%%%%%%%%%%%%%%%%%%%%%%%%%%%%%%%%%%%

%%%%%%%%%%%%%%%%%%%%%%%%%%%%%%%%%%%%%%%%%%%%%%%%%%%%%%%%%%%%%%%%%%%%%%%%%%%%%%%%%%%%%%
%
%
%
%%%%%%%%%%%%%%%%%%%%%%%%%%%%%%%%%%%%%%%%%%%%%%%%%%%%%%%%%%%%%%%%%%%%%%%%%%%%%%%%%%%%%%
\begin{table*}[htb]
\caption{
The total cross section of the radiative recombination of an electron
into the $(1s)^2$, $1s2s$, $1s2p_{1/2}$ and $1s2p_{3/2}$ states of the
initially hydrogen-like uranium, for different values of the projectile energy
$E$. $\sigma^{(0)}$ is the zeroth-order cross
section. $\sigma_{\rm scr}$ is the screening correction,
$\sigma^{(1)}_{\rm corr}$ is the correction due to the electron correlation on the
bound electron state, and $\sigma^{(1)}_{\rm DR}$ is the correction due to the
dielectronic recombination. All corrections are given in units per cent of
$\sigma^{(0)}$.
\label{tab:1}}
\begin{ruledtabular}
  \begin{tabular}{c........}
 \multicolumn{1}{c}{$E$} &
 \multicolumn{1}{c}{$\sigma^{(0)}$} &
 \multicolumn{1}{c}{$\sigma_{\rm scr}$} &
 \multicolumn{1}{c}{$\sigma^{(1)}_{\rm corr}$} &
 \multicolumn{1}{c}{$\sigma^{(1)}_{\rm DR}$} &
 \multicolumn{1}{c}{$\sigma^{(0)}$} &
 \multicolumn{1}{c}{$\sigma_{\rm scr}$} &
 \multicolumn{1}{c}{$\sigma^{(1)}_{\rm corr}$} &
 \multicolumn{1}{c}{$\sigma^{(1)}_{\rm DR}$}
\\[0.1cm]
 \multicolumn{1}{c}{[MeV$\!/$u]} &
 \multicolumn{1}{c}{[barn]} &
 \multicolumn{1}{c}{[\%]} &
 \multicolumn{1}{c}{[\%]} &
 \multicolumn{1}{c}{[\%]} &
 \multicolumn{1}{c}{[barn]} &
 \multicolumn{1}{c}{[\%]} &
 \multicolumn{1}{c}{[\%]} &
 \multicolumn{1}{c}{[\%]}
\\[0.1cm]
\hline
\\[-0.1cm]
        &  \multicolumn{4}{c}{$(1s)^2$ state}   &  \multicolumn{4}{c}{$1s2s$ state}
\\[0.2cm]
  1 &  1.588x\times 10^4 & -0.x850 &  0.x138 & -0.x703    &  5.080x\times 10^3 & -1.x997 & -0.x232 &  0.x387   \\
  2 &  7.917x\times 10^3 & -0.x851 &  0.x140 & -0.x702    &  2.539x\times 10^3 & -1.x981 & -0.x228 &  0.x390  \\ 
  5 &  3.142x\times 10^3 & -0.x852 &  0.x146 & -0.x702    &  1.013x\times 10^3 & -1.x936 & -0.x217 &  0.x396   \\
 10 &  1.550x\times 10^3 & -0.x854 &  0.x156 & -0.x701    &  5.040x\times 10^2 & -1.x874 & -0.x201 &  0.x408   \\
 25 &  5.967x\times 10^2 & -0.x864 &  0.x182 & -0.x696    &  1.968x\times 10^2 & -1.x755 & -0.x164 &  0.x444   \\
 50 &  2.806x\times 10^2 & -0.x888 &  0.x217 & -0.x683    &  9.315x\times 10   & -1.x673 & -0.x124 &  0.x524   \\
 75 &  1.765x\times 10^2 & -0.x917 &  0.x245 & -0.x668    &  5.841x\times 10   & -1.x652 & -0.x101 &  0.x669   \\
100 &  1.254x\times 10^2 & -0.x949 &  0.x268 & -0.x650    &  4.118x\times 10   & -1.x655 & -0.x086 &  1.x141   \\
125 &  9.524x\times 10   & -0.x981 &  0.x286 & -0.x634    &  3.101x\times 10   & -1.x670 & -0.x076 & -6.x124   \\
150 &  7.559x\times 10   & -1.x013 &  0.x302 & -0.x617    &  2.438x\times 10   & -1.x690 & -0.x069 &  1.x635   \\
175 &  6.188x\times 10   & -1.x043 &  0.x314 & -0.x601    &  1.977x\times 10   & -1.x712 & -0.x065 &  0.x443   \\
200 &  5.184x\times 10   & -1.x072 &  0.x325 & -0.x587    &  1.642x\times 10   & -1.x735 & -0.x062 & -0.x044   \\
250 &  3.827x\times 10   & -1.x125 &  0.x341 & -0.x562    &  1.192x\times 10   & -1.x778 & -0.x060 & -0.x034   \\
300 &  2.968x\times 10   & -1.x170 &  0.x353 & -0.x542    &  9.102x            & -1.x817 & -0.x060 & -0.x017   \\
400 &  1.966x\times 10   & -1.x242 &  0.x369 & -0.x515    &  5.879x            & -1.x880 & -0.x063 &  0.x015   \\
500 &  1.419x\times 10   & -1.x292 &  0.x378 & -0.x499    &  4.159x            & -1.x926 & -0.x068 &  0.x034   \\
600 &  1.084x\times 10   & -1.x327 &  0.x383 & -0.x492    &  3.127x            & -1.x958 & -0.x074 &  0.x043   \\
700 &  8.621x            & -1.x350 &  0.x387 & -0.x495    &  2.456x            & -1.x980 & -0.x079 &  0.x045   \\
\\
        &  \multicolumn{4}{c}{$1s2p_{1/2}$ state}   &        \multicolumn{4}{c}{$1s2p_{3/2}$ state}
\\[0.2cm]
  1 &  7.227x\times 10^3 & -2.x479 & -0.x084 & -0.x045    &  9.951x\times 10^3 & -1.x998 &  0.x058 & -0.x014   \\
  2 &  3.574x\times 10^3 & -2.x507 & -0.x087 & -0.x046    &  4.888x\times 10^3 & -2.x034 &  0.x061 & -0.x015   \\
  5 &  1.384x\times 10^3 & -2.x589 & -0.x096 & -0.x051    &  1.856x\times 10^3 & -2.x140 &  0.x067 & -0.x019   \\
 10 &  6.568x\times 10^2 & -2.x715 & -0.x111 & -0.x060    &  8.546x\times 10^2 & -2.x299 &  0.x076 & -0.x025   \\
 25 &  2.272x\times 10^2 & -3.x034 & -0.x147 & -0.x090    &  2.737x\times 10^2 & -2.x682 &  0.x098 & -0.x049   \\
 50 &  9.209x\times 10   & -3.x422 & -0.x195 & -0.x160    &  1.006x\times 10^2 & -3.x119 &  0.x123 & -0.x106   \\
 75 &  5.138x\times 10   & -3.x695 & -0.x232 & -0.x272    &  5.205x\times 10   & -3.x415 &  0.x140 & -0.x205   \\
100 &  3.304x\times 10   & -3.x897 & -0.x261 & -0.x520    &  3.148x\times 10   & -3.x630 &  0.x152 & -0.x449   \\
125 &  2.308x\times 10   & -4.x053 & -0.x286 & 14.x715    &  2.090x\times 10   & -3.x793 &  0.x160 & 11.x296   \\
150 &  1.705x\times 10   & -4.x176 & -0.x307 & -0.x666    &  1.477x\times 10   & -3.x922 &  0.x166 &  0.x127   \\
175 &  1.312x\times 10   & -4.x276 & -0.x326 & -0.x222    &  1.093x\times 10   & -4.x026 &  0.x171 &  0.x352   \\
200 &  1.040x\times 10   & -4.x358 & -0.x342 & -0.x047    &  8.375x            & -4.x113 &  0.x174 &  0.x248   \\
250 &  7.006x            & -4.x487 & -0.x369 &  0.x021    &  5.319x            & -4.x246 &  0.x179 &  0.x297   \\
300 &  5.041x            & -4.x581 & -0.x392 &  0.x055    &  3.646x            & -4.x343 &  0.x181 &  0.x306   \\
400 &  2.977x            & -4.x712 & -0.x427 &  0.x071    &  1.995x            & -4.x473 &  0.x182 &  0.x275   \\
500 &  1.972x            & -4.x796 & -0.x455 &  0.x055    &  1.248x            & -4.x551 &  0.x181 &  0.x222   \\
600 &  1.410x            & -4.x854 & -0.x478 &  0.x027    &  0.853x            & -4.x597 &  0.x179 &  0.x166   \\
700 &  1.063x            & -4.x895 & -0.x499 & -0.x004    &  0.620x            & -4.x623 &  0.x176 &  0.x111   \\
  \end{tabular}
\end{ruledtabular}
\end{table*}

%%%%%%%%%%%%%%%%%%%%%%%%%%%%%%%%%%%%%%%%%%%%%%%%%%%%%%%%%%%%%%%%%%%%%%%%%%%%%%%%%%%%%%
%
%
%
%%%%%%%%%%%%%%%%%%%%%%%%%%%%%%%%%%%%%%%%%%%%%%%%%%%%%%%%%%%%%%%%%%%%%%%%%%%%%%%%%%%%%%
\begin{table*}[htb]
\caption{
The total cross section of the radiative recombination of an electron
into the $(1s)^2$ and $1s2s$ states of the
initially hydrogen-like tin ($Z=50$), for different values of the projectile energy
$E$. Notations are the same as in Table~\ref{tab:1}.
\label{tab:2}}
\begin{ruledtabular}
  \begin{tabular}{c........}
 \multicolumn{1}{c}{$E$} &
 \multicolumn{1}{c}{$\sigma^{(0)}$} &
 \multicolumn{1}{c}{$\sigma_{\rm scr}$} &
 \multicolumn{1}{c}{$\sigma^{(1)}_{\rm corr}$} &
 \multicolumn{1}{c}{$\sigma^{(1)}_{\rm DR}$} &
 \multicolumn{1}{c}{$\sigma^{(0)}$} &
 \multicolumn{1}{c}{$\sigma_{\rm scr}$} &
 \multicolumn{1}{c}{$\sigma^{(1)}_{\rm corr}$} &
 \multicolumn{1}{c}{$\sigma^{(1)}_{\rm DR}$}
\\[0.1cm]
 \multicolumn{1}{c}{[MeV$\!/$u]} &
 \multicolumn{1}{c}{[barn]} &
 \multicolumn{1}{c}{[\%]} &
 \multicolumn{1}{c}{[\%]} &
 \multicolumn{1}{c}{[\%]} &
 \multicolumn{1}{c}{[barn]} &
 \multicolumn{1}{c}{[\%]} &
 \multicolumn{1}{c}{[\%]} &
 \multicolumn{1}{c}{[\%]}
\\[0.1cm]
\hline
\\[-0.1cm]
        &  \multicolumn{4}{c}{$(1s)^2$ state}   &  \multicolumn{4}{c}{$1s2s$ state}
\\[0.2cm]
  1 &  5.036x\times 10^3 & -1.x373 &  0.x399 & -0.x981    &  1.528x\times 10^3 & -3.x407 & -0.x271 &  0.x405   \\
  2 &  2.492x\times 10^3 & -1.x370 &  0.x419 & -0.x982    &  7.598x\times 10^2 & -3.x345 & -0.x244 &  0.x399   \\
  5 &  9.669x\times 10^2 & -1.x362 &  0.x473 & -0.x983    &  2.976x\times 10^2 & -3.x218 & -0.x174 &  0.x388   \\
 10 &  4.603x\times 10^2 & -1.x356 &  0.x552 & -0.x984    &  1.425x\times 10^2 & -3.x116 & -0.x090 &  0.x386   \\
 25 &  1.605x\times 10^2 & -1.x364 &  0.x721 & -0.x970    &  4.931x\times 10   & -3.x071 &  0.x046 &  0.x532   \\
 50 &  6.568x\times 10   & -1.x408 &  0.x873 & -0.x929    &  1.965x\times 10   & -3.x133 &  0.x129 &  0.x222   \\
 75 &  3.684x\times 10   & -1.x455 &  0.x947 & -0.x887    &  1.078x\times 10   & -3.x187 &  0.x152 &  0.x174   \\
100 &  2.375x\times 10   & -1.x494 &  0.x985 & -0.x852    &  6.835x            & -3.x223 &  0.x153 &  0.x133   \\
125 &  1.661x\times 10   & -1.x526 &  1.x003 & -0.x813    &  4.718x            & -3.x247 &  0.x144 &  0.x098   \\
150 &  1.226x\times 10   & -1.x552 &  1.x010 & -0.x804    &  3.447x            & -3.x265 &  0.x131 &  0.x069   \\
175 &  9.419x            & -1.x575 &  1.x011 & -0.x796    &  2.626x            & -3.x277 &  0.x117 &  0.x043   \\
200 &  7.456x            & -1.x593 &  1.x009 & -0.x761    &  2.064x            & -3.x286 &  0.x102 &  0.x022   \\
250 &  4.996x            & -1.x624 &  0.x999 & -0.x740    &  1.367x            & -3.x300 &  0.x073 & -0.x011   \\
300 &  3.575x            & -1.x647 &  0.x986 & -0.x677    &  0.970x            & -3.x308 &  0.x047 & -0.x035   \\
400 &  2.088x            & -1.x680 &  0.x959 & -0.x618    &  0.559x            & -3.x317 &  0.x003 & -0.x063   \\
500 &  1.372x            & -1.x699 &  0.x935 & -0.x601    &  0.364x            & -3.x319 & -0.x031 & -0.x080   \\
600 &  0.974x            & -1.x711 &  0.x914 & -0.x581    &  0.257x            & -3.x316 & -0.x059 & -0.x089   \\
700 &  0.732x            & -1.x716 &  0.x896 & -0.x554    &  0.192x            & -3.x310 & -0.x083 & -0.x094   \\
  \end{tabular}
\end{ruledtabular}
\end{table*}

%%%%%%%%%%%%%%%%%%%%%%%%%%%%%%%%%%%%%%%%%%%%%%%%%%%%%%%%%%%%%%%%%%%%%%%%%%%%%%%%%%
%
%%%%%%%%%%%%%%%%%%%%%%%%%%%%%%%%%%%%%%%%%%%%%%%%%%%%%%%%%%%%%%%%%%%%%%%%%%%%%%%%%%
\section{Results and discussion}

The calculational results for the total cross section of the RR of an electron with
an (initially) hydrogen-like uranium are presented in
Table~\ref{tab:1} for the capture into the $(1s)^2$, 
$1s2s$, $1s2p_{1/2}$, and $1s2p_{3/2}$ states. 
$\sigma^{(0)}$ is the zeroth-order cross section.
It is calculated with the energy of the emitted photon that includes 
all known {\em one-electron} corrections to the energy of the final state, 
i.e., $\omega = \vare-m-\vare_{{\rm io},H}$, where $\vare_{{\rm
  io},H}$ is the ionization energy of the hydrogen-like ion.
$\sigma_{\rm scr}$ is the correction due to the screening of the nuclear
charge by the core electron. It was obtained by re-evaluating the
zeroth-order cross section with the wave functions calculated
in the presence of the screening potential. The energy of the
emitted photon includes all known corrections to the energy of the final
state \cite{artemyev:05:pra}, i.e., $\omega = \vare-m-\vare_{{\rm io}}$, where $\vare_{{\rm
  io}}$ is the ionization energy of the helium-like ion.
$\sigma^{(1)}_{\rm corr}$
is the correlation correction induced by $\tau^{(1,5)}$.
$\sigma^{(1)}_{\rm DR}$ is the off-resonant
DR correction induced by $\tau^{(1,3)}$, $\tau^{(1,4)}$,
$\tau^{(1,6)}$, $\tau^{(1,7)}$, and $\tau^{(1,8)}$. For the recombination into the
excited states, $\sigma^{(1)}_{\rm DR}$ contains a
series of the DR resonance peaks in the region of
projectile energies $E = 110-190$ MeV/u. The
behaviour of $\sigma^{(1)}_{\rm DR}$ in the vicinity of the peaks is
shown in Fig.~\ref{fig:2}. In the case of recombination into the ground state,
$\sigma^{(1)}_{\rm DR}$ does not have any resonances.

Our calculation shows that the effect of the screening of the nuclear charge
generally grows for larger projectile energies and the capture into
higher excited states, approaching the limit of the complete screening (i.e., 
the case of the capture
by a bare nucleus with the $Z-1$ charge). 
The effect of the off-resonant DR mechanism
is the strongest for the capture into the ground state and for low projectile
energies. In this case, the DR contribution is of 
the similar size as the contribution of the screening effect.
We conclude that for the capture into the ground state, 
the electron-electron interaction needs to be accounted for rigorously and with
inclusion of the off-resonant DR mechanism. 
Results obtained by an effective one-electron theory or by standard
many-body approaches such as the Dirac-Fock method provide only an
order-of-magnitude estimate of the two-electron effect in this case.
However, for the
recombination into the excited states and the projectile energy beyond
the DR resonance threshold, the DR correction is much smaller
than the screening contribution and can be neglected for most
practical purposes.
For the projectile energies below the threshold, the off-resonant DR mechanism can
be important in the vicinity of the peaks, even at relatively large
distances from the region of resonance.

In order to illustrate the dependence of the effects studied on the nuclear
charge number $Z$, Table~\ref{tab:2} presents the calculational results for the
recombination into the $(1s)^2$ and $1s2s$ states of the initially hydrogen-like
tin ($Z=50$). We observe that the relative contribution of the screening
effect is roughly proportional to $1/Z$, as could be expected. It is
remarkable that the electron correlation correction, which plays only a minor
role for uranium, becomes important for tin in the case of capture into the
ground state. The relative contribution of the off-resonant DR
mechanism is slightly larger for tin than for uranium, but, in comparison to
the screening effect, the DR correction becomes somewhat
less significant for lighter ions.

In Fig.~\ref{fig:3} we present the results for the differential cross section
for the case of the capture into the ground state of 
uranium, for two values of the projectile energy
$E = 50$ and 300 MeV/u, which are typical for the ESR storage ring at GSI. 
The differential cross section is calculated in the laboratory frame, in which
the initially free electron is at rest. We observe that the screening and the DR
contributions have different dependence on the observation angle. For the zero
angle, they are of the opposite sign and
significantly cancel each other, whereas for larger angles these two effects 
amplify each other.

One of the motivations of the present study was a deviation 
from predictions of one-electron theory reported in the
experimental investigation of RR into a hydrogen-like uranium
at very small projectile velocities \cite{reuschl:08}. 
An effect of about 10\% was observed in the experiment, whereas
a much smaller contribution on the level of 1-2\%  
was expected from theory \cite{fritzsche:unpub}.

Our {\em ab initio} calculation demonstrates that the
electron-electron interaction affects the RR cross section
on the level of about 2\% for the projectile energies of several MeV/u,  
which agrees with previous estimates. For smaller projectile energies,
the cross section is well described by the asymptotic behaviour $E\,\sigma(E) =
\mbox{\rm const}$,  and the relative values of all corrections stay constant.
So, our calculation cannot explain the large two-electron effect observed in
Ref.~\cite{reuschl:08}. We note, however, that the quantities actually 
measured in this experiment were not the cross sections but the recombination
rates. A consistent interpretation of the experimental results
requires a careful consideration of the recombination rates under the
experimental conditions. Such a calculation in underway and will be reported 
elsewhere.

%%%%%%%%%%%%%%%%%%%%%%%%%%%%%%%%%%%%%%%%%%%%%%%%%%%%%%%%%%%%%%%%%%%%%%%%%%%%%%%%%%
\section{Summary}

We have performed an investigation of the radiative
recombination of an electron with an (initially) hydrogen-like ion. The
electron-electron interaction was treated rigorously to the first order in
the parameter $1/Z$ and within the screening-potential approximation to the
higher orders in $1/Z$. The contribution of the off-resonant dielectronic
recombination was studied in detail. It was demonstrated that this mechanism
contributes significantly to the total effect of the electron-electron
interaction in the case of recombination into the ground state. For the
recombination into the excited states, it is significant
in the vicinity of the resonance peaks but becomes small for the projectile
energies beyond the resonant dielectronic-recombination threshold.

The work reported in this paper was supported by the Helmholtz Gemeinschaft
(Nachwuchsgruppe VH-NG-421).

%%%%%%%%%%%%%%%%%%%%%%%%%%%%%%%%%%%%%%%%%%%%%%%%%%%%%%%%%%%%%%%%%%%%%%%%%%%%%%%%%%%%%%
%
%
%
%%%%%%%%%%%%%%%%%%%%%%%%%%%%%%%%%%%%%%%%%%%%%%%%%%%%%%%%%%%%%%%%%%%%%%%%%%%%%%%%%%%%%%
\appendix*

\section{Calculational formulas}

The spherical-wave expansion of the Dirac
wave function of an incident electron
with a fixed asymptotic momentum is \cite{eichler:95:book}
\begin{align}
|\bm{p}\mu_s\bigr>  = 4\pi\, \sum_{\kappa\mu} i^l\, e^{i\Delta_{\kappa}}\,C^{j\mu}_{l m_l\, \frac12
  \mu_s}\, Y_{lm_l}^*(\hp)\,   |\vare \kappa \mu\bigr>
\,,
\end{align}
where $j = |\kappa|-1/2$, $l = |\kappa+1/2|-1/2$, $\Delta_{\kappa}$ is the
phase shift, and $|\vare \kappa \mu\bigr>$ is the continuum Dirac
wave function with the relativistic angular quantum number $\kappa$ and the
angular momentum projection $\mu$, normalized on the energy scale.
After the integration over the angular variables (see
Ref.~\cite{yerokhin:00:recpra} for details), the result for the
zeroth-order amplitude is given by
\begin{align} \label{eq5}
\tau^{(0)}_{\mu_s,\mu_v}(\hp) &\,  =
4\pi\, \sum_{\kappa} i^l\, e^{i\Delta_k}\,C^{j\mu}_{l m_l\, \frac12
  \mu_s}\, Y_{lm_l}^*(\hp)\,
  \sum_{JL} i^{-1-L}\,
 \nonumber \\ & \times
\sqrt{2L+1}\,
     C^{JM}_{L0\,1\lambda}\, (-1)^{j-\mu}\,C^{JM}_{j_v\mu_v\,j-\mu}\, P_{JL}(\omega,v\vare)
\,,
\end{align}
where the radial integrals $P_{JL}$ are defined as
\begin{align}
P_{JL}(\omega,ab) =& \, \int_0^{\infty}dx\,x^2\,j_L(\omega x)\,\bigl[
  g_b(x)\,f_a(x)\,S_{JL}(\kappa_b,-\kappa_a)
 \nonumber \\ &
 -f_b(x)\,g_a(x)\,S_{JL}(-\kappa_b,\kappa_a)\bigr]\,.
\end{align}
The angular coefficients $S_{JL}(\kappa_1,\kappa_2)$ are
given, e.g., by Eqs.~(C7)-(C9) of Ref.~\cite{yerokhin:99:pra}.
The momentum projections $\mu$, $M$, and $m_l$ in Eq.~(\ref{eq5}) are fixed
by the selection rules of Clebsch-Gordan coefficients.
$\lambda = \pm 1$ corresponds to the circular polarization of the emitted photon.
(The cross section does not depend on the sign of $\lambda$.)

The one-photon exchange corrections to the transition amplitude $\tau^{(1,i)}$
can be expressed in the form similar to that for the zeroth-order amplitude,
with the radial integrals $P_{JL}$ substituted by
their generalizations $\mathcal{F}^{(1,i)}_{JL}$. The results for the functions
$\mathcal{F}^{(1,i)}_{JL}$ are
\begin{align} \label{F3}
{\mathcal F}^{(1,3)}_{JL} = &\
\alpha \sum_{n} \frac{P_{JL}(\omega,na)}{\vare_a+\vare_v-\vare-\vare_n}\,
 \nonumber \\ & \times
   \frac{(-1)^{J+j_a-j_n}}{2J+1}\, R_{J}(\vare-\vare_v,va\vare n)\,,
\end{align}
\begin{align}
{\mathcal F}^{(1,4)}_{JL} = &\
\alpha \sum_{n} \frac{P_{JL}(\omega,an)}{\vare_a-\vare_v+\vare-\vare_n}\,
 \nonumber \\ & \times
   \frac{(-1)^{J+j_a-j_n}}{2J+1}\, R_{J}(\vare-\vare_v,vn\vare a)\,,
\end{align}
\begin{align} \label{F5}
{\mathcal F}^{(1,5)}_{JL} = &\
\alpha \sum_{{n\ne v}\atop{\kappa_n = \kappa_v}} \frac{P_{JL}(\omega,n\vare)}{\vare_v-\vare_n}\,
  \sum_{L_0}
     \frac{(-1)^{j_a+j_v+L_0}}{2j_v+1}\,
 \nonumber \\ & \times
R_{L_0}(\vare_v-\vare_a,avna)
+\frac{\alpha}2\, P_{JL}(\omega,v\vare)\,
 \nonumber \\ & \times
  \sum_{L_0}
    \frac{(-1)^{j_a+j_v+L_0}}{2j_v+1}\,R^{\,\prime}_{L_0}(\vare_v-\vare_a,avva)\,,
\end{align}
\begin{align}
{\mathcal F}^{(1,6)}_{JL} = &\
\alpha \sum_{{n}\atop{\kappa_n = \kappa}} \frac{P_{JL}(\omega,vn)}{\vare-\vare_n}\,
 \nonumber \\ & \times
  \sum_{L_0}
    \frac{(-1)^{j_a+j+L_0}}{2j+1}\, R_{L_0}(\vare-\vare_a,naa\vare)\,,
 \label{F6}
\end{align}
\begin{align} \label{F7}
{\mathcal F}^{(1,7)}_{JL} = &\
\alpha \sum_{n} \frac{P_{JL}(\omega,na)}{\vare_a+\vare_v-\vare-\vare_n}\,
  \sum_{L_0}
    (-1)^{j_a-j_n+J}
 \nonumber \\ & \times
\SixJ{j_v}{j}{J}{j_a}{j_n}{L_0}\, R_{L_0}(\vare-\vare_a,av\vare n)\,,
\end{align}
\begin{align} \label{F8}
{\mathcal F}^{(1,8)}_{JL} = &\
\alpha \sum_{n} \frac{P_{JL}(\omega,an)}{\vare_a-\vare_v+\vare-\vare_n}\,
  \sum_{L_0}
    (-1)^{j_a-j_n+J}
 \nonumber \\ & \times
\SixJ{j}{j_v}{J}{j_a}{j_n}{L_0}\, R_{L_0}(\vare_v-\vare_a,vna\vare )\,,
\end{align}
where $R_L$ is the relativistic generalization of the Slater integral
(see Appendix C of Ref.~\cite{yerokhin:99:pra}). The prime of $R_L$ in
Eq.~(\ref{F5}) denotes the derivative with respect to the energy argument,
$R^{\,\prime}_L(\vare,abcd) = \left. d/(d\omega)\,R_L(\omega,abcd)
\right|_{\omega = \vare}$.

%\bibliographystyle{/home/yerokhin/papers/bibtex/phaip30}
%\bibliography{/home/yerokhin/papers/bibtex/hfst}

\begin{thebibliography}{10}

\bibitem{spies:92}
W.~Spies, A.~M\"uller, J.~Linkemann, A.~Frank, M.~Wagner, C.~Kozhuharov,
  B.~Franzke, K.~Beckert, F.~Bosch, H.~Eickhoff, M.~Jung, O.~Klepper,
  W.~K\"onig, P.~H. Mokler, R.~Moshammer, F.~Nolden, U.~Schaaf, P.~Sp\"adtke,
  M.~Steck, P.~Zimmerer, N.~Gr\"un, W.~Scheid, M.~S. Pindzola, and N.~R.
  Badnell,
\newblock Phys. Rev. Lett. {\bf 69}, 2768 (1992).

\bibitem{zimmermann:97}
M.~Zimmermann, N.~Gr\"un, and W.~Scheid,
\newblock J. Phys. B {\bf 30}, 5259 (1997).

\bibitem{tokman:02}
M.~Tokman, N.~Ekl\"ow, P.~Glans, E.~Lindroth, R.~Schuch, G.~Gwinner,
  D.~Schwalm, A.~Wolf, A.~Hoffknecht, A.~M\"uller, and S.~Schippers,
\newblock Phys. Rev. A {\bf 66}, 012703 (2002).

\bibitem{mohamed:02}
T.~Mohamed, D.~Nikoli\ifmmode~\acute{c}\else \'{c}\fi{}, E.~Lindroth,
  S.~Madzunkov, M.~Fogle, M.~Tokman, and R.~Schuch,
\newblock Phys. Rev. A {\bf 66}, 022719 (2002).

\bibitem{stoehlker:94}
T.~St\"ohlker, H.~Geissel, H.~Irnich, T.~Kandler, C.~Kozhuharov, P.~H. Mokler,
  G.~M\"unzenberg, F.~Nickel, C.~Scheidenberger, T.~Suzuki, M.~Kucharski,
  A.~Warczak, P.~Rymuza, Z.~Stachura, A.~Kriessbach, D.~Dauvergne, B.~Dunford,
  J.~Eichler, A.~Ichihara, and T.~Shirai,
\newblock Phys. Rev. Lett. {\bf 73}, 3520 (1994).

\bibitem{stoehlker:97:prl}
T.~St\"ohlker, F.~Bosch, A.~Gallus, C.~Kozhuharov, G.~Menzel, P.~H. Mokler,
  H.~T. Prinz, J.~Eichler, A.~Ichihara, T.~Shirai, R.~W. Dunford,
  T.~Ludziejewski, P.~Rymuza, Z.~Stachura, P.~Swiat, and A.~Warczak,
\newblock Phys. Rev. Lett. {\bf 79}, 3270 (1997).

\bibitem{stoehlker:99:prl}
T.~St\"ohlker, T.~Ludziejewski, F.~Bosch, R.~W. Dunford, C.~Kozhuharov, P.~H.
  Mokler, H.~F. Beyer, O.~Brinzanescu, B.~Franzke, J.~Eichler, A.~Griegal,
  S.~Hagmann, A.~Ichihara, A.~Kr\"amer, J.~Lekki, D.~Liesen, F.~Nolden,
  H.~Reich, P.~Rymuza, Z.~Stachura, M.~Steck, P.~Swiat, and A.~Warczak,
\newblock Phys. Rev. Lett. {\bf 82}, 3232  (1999),
\newblock [(E) {\em ibid}, {\bf 84}, 1360 (2000)].

\bibitem{reuschl:08}
R.~Reuschl, A.~Gumberidze, C.~Kozhuharov, U.~Spillmann, S.~Tashenov,
  T.~St\"ohlker, and J.~Eichler,
\newblock Phys. Rev. A {\bf 77}, 032701 (2008).

\bibitem{fritzsche:05}
S.~Fritzsche, A.~Surzhykov, and T.~St\"ohlker,
\newblock Phys. Rev. A {\bf 72}, 012704 (2005).

\bibitem{trzhaskovskaya:03}
M.~B. Trzhaskovskaya and V.~K. Nikulin,
\newblock Optics and Spectroscopy {\bf 95}, 537 (2003)
\newblock [Optika i Spektroskopiya {\bf 95}, 580 (2003)].

\bibitem{korol:06}
A.~V. Korol, G.~F. Gribakin, and F.~J. Currell,
\newblock Phys. Rev. Lett. {\bf 97}, 223201 (2006).

\bibitem{korol:04}
A.~V. Korol, F.~J. Currell, and G.~F. Gribakin,
\newblock J. Phys. B {\bf 37}, 2411 (2004).

\bibitem{yerokhin:00:recpra}
V.~A. Yerokhin, V.~M. Shabaev, T.~Beier, and J.~Eichler,
\newblock Phys. Rev. A {\bf 62}, 042712 (2000).

\bibitem{eichler:95:book}
J.~Eichler and W.~Meyerhof,
\newblock {\em Relativistic Atomic Collisions}
\newblock (Academic Press, San Diego, 1995).

\bibitem{shabaev:02:rep}
V.~M. Shabaev,
\newblock Phys. Rep. {\bf 356}, 119  (2002).

\bibitem{salvat:95:cpc}
F.~Salvat, J.~M. Fern\'{a}ndez-Varea, and W.~{Williamson Jr.},
\newblock Comput. Phys. Commun. {\bf 90}, 151  (1995).

\bibitem{shabaev:04:DKB}
V.~M. Shabaev, I.~I. Tupitsyn, V.~A. Yerokhin, G.~Plunien, and G.~Soff,
\newblock Phys. Rev. Lett. {\bf 93}, 130405 (2004).

\bibitem{mohr:98}
P.~J. Mohr, G.~Plunien, and G.~Soff,
\newblock Phys. Rep. {\bf 293}, 227  (1998).

\bibitem{artemyev:05:pra}
A.~N. Artemyev, V.~M. Shabaev, V.~A. Yerokhin, G.~Plunien, and G.~Soff,
\newblock Phys. Rev. A {\bf 71}, 062104 (2005).

\bibitem{fritzsche:unpub}
S.~Fritzsche and A.~Surzhykov,
\newblock unpublished.

\bibitem{yerokhin:99:pra}
V.~A. Yerokhin and V.~M. Shabaev,
\newblock Phys. Rev. A {\bf 60}, 800  (1999).

\end{thebibliography}

\end{document}